\documentclass[aps,prl,nobibnotes,nofootinbib,showpacs,reprint]{revtex4-1}

\usepackage{graphics}
\usepackage{amsmath}

\newcommand{\Fig}[1]{FIG.~\ref{#1}}
\newcommand{\Eq}[1]{Eq.~(\ref{#1})}
\newcommand{\infint}{\int_{-\infty}^{+\infty}}

\newcommand{\vect}[1]{\mathbf{#1}}

\begin{document}

\title{Measurement of the Ultrafast Temporal Response of a Plasmonic Antenna}

\author{Maria~Becker}
\email{email: mgbecker3@gmail.com}

\author{Wayne~Cheng-Wei~Huang}
\email{email: u910134@alumni.nthu.edu.tw}

\author{Herman~Batelaan}
\email{email: hbatelaan@unl.edu}

\affiliation{Department of Physics and Astronomy, University of Nebraska-Lincoln, Lincoln, Nebraska 68588, USA}

\author{Elizabeth~J.~Smythe}
\email{Current affiliation: Schlumberger-Doll Research in Cambridge, MA.}

\author{Federico~Capasso}

\affiliation{Harvard School of Engineering and Applied Science, Harvard University, Cambridge, Massachusetts 02138, USA}

\begin{abstract}
	We report a measurement on the temporal response of a plasmonic antenna at the femtosecond time scale. The antenna consists of a square array of nanometer-size gold rods. We find that the far-field dispersion of light reflected from the plasmonic antenna is less than that of a $1.2 mm$ thick glass slide. Assuming a simple oscillating dipole model this implies that the near-field of the antenna may be used as an electron switch that responds faster than $20 fs$. Alternatively, ultrafast electron diffraction may be used to investigate the near-field dynamics of the plasmonic antenna.
\end{abstract}


\maketitle

Light incident on a metallic material is strongly attenuated at the surface. For metallic films of nanometer-scale thickness, all electrons in the material can interact with light at optical frequencies. When this thin metallic film is patterned with nanometer scale boundaries to form a metallic nanoparticle, electron resonance behavior can occur. This phenomenon can take place when the dominant spatial mode of the charge oscillation, i.e.\ the dipole oscillation mode, corresponds to the boundary shape of the particle \cite{Aizpurua}. Consequently, metallic nanoparticles that are made of the same material but have different shapes reflect different colors under white light illumination \cite{Murray,PhysicsToday}.

As the size of metallic particles shrinks down to sub-wavelength dimensions, a strong near-field can arise at the poles of the oscillating dipole. The magnitude of this near field can exceed ten times that of the incident field which drives the particle, a phenomena known as near-field enhancement \cite{Capasso:enhancement, Kelly}. The near-field enhancement can be even greater if several metallic particles are positioned close to each other \cite{Capasso:enhancement}. Additionally, the resonant response of a particle assembly can be tuned through variation of the inter-particle distances \cite{Capasso:resonance,Rechberger}. Thus, an array of metallic nanoparticles can be used as an antenna, known as a plasmonic antenna \cite{Capasso:resonance}. The periodic structure and strong near-field enhancement of a plasmonic antenna may enable it to act as an electron diffraction grating. In order to utilize a plasmonic antenna as an electron diffraction/deflection device in the ultrafast regime, it is important to know how a plasmonic antenna responds to ultrafast light. In this Letter, we investigate the temporal response of a plasmonic antenna upon excitation by a femtosecond light pulse.

The temporal response of regular arrays of nanoparticles have been previously studied with methods sensitive to a variety of nanoparticle shapes \cite{Lamprecht:1,Lamprecht:2,Lamprecht:3,Lamprecht:4}. The results from autocorrelation measurements of transmitted second- and third- harmonic generation signals indicated a fast relaxation time of $6fs$. The arrays used had periodicities and gapsizes of hundreds of nanometers. In this study, we use a reflective cross-correlation method, which does not require higher-harmonic generation. We also propose that this arrangement could be used for electron manipulation. Periodic structures with a period of about 100 nm and gaps of tens of nanometers are of particular interest for electron diffraction. The former provides practical diffraction angles, while the latter yields an intensity enhancement that may be suited for femtosecond electron switching. In view of our proposed application, the main objective of the present work is to establish that the enhanced near-fields in the small gaps between elements of the array studied do not significantly lengthen the temporal response of the reflected pulse.

The near-field of nanoscale structures has been probed time-independently with tip-enhanced electron emission microscopy \cite{Ropers} and time-dependently with photon-induced near-field microscopy \cite{Barwick}. As pointed out previously in \cite{Barwick}, these techniques may also be used to investigate plasmonic antennas. The simple all-optical technique presented in this letter may be useful for selecting plasmonic antennas with an interesting response in the far-field that warrant a more involved study with pulsed electron techniques.

When excited by a light pulse, the plasmonic antenna generates a strong localized near-field. The far-field is radiated as a reflected pulse. For an array of dipole radiators, the relation between the antenna's near- and far-fields can be deduced by inspecting the fields of each single oscillating dipole, $p(t) = p_{\omega}e^{-i(\omega t+\phi_{\omega})}$ \cite{Jackson},
\begin{equation}\label{dipolefield}
	\begin{split}
	¥	\vect{E}_{dip,\omega}(\vect{r},t) &= \frac{1}{4\pi\epsilon_{0}} \left\{ \frac{k^2}{r}(\hat{\vect{r}} \times \hat{\vect{z}}) \times \hat{\vect{r}} \right. 		\\
	&\quad + \left. \left( \frac{1}{r^3} - \frac{ik}{r^2} \right) [ 3( \hat{\vect{r}} \cdot \hat{\vect{z}} )\hat{\vect{r}}  - \hat{\vect{z}} ] \right\}p_{\omega}e^{-i(\omega t+\phi_{\omega})}e^{ikr} 	\\
	&= E_{far}(\vect{r}) p_{\omega} \omega^2 e^{i\phi_{\omega}} e^{-i\omega(t-r/c)} 		\\
	&\quad + E_{mid}(\vect{r}) p_{\omega} \omega e^{i\pi/2} e^{i\phi_{\omega}} e^{-i\omega(t-r/c)} 	\\
	&\quad + E_{near}(\vect{r}) p_{\omega} e^{i\phi_{\omega}} e^{-i\omega(t-r/c)},
	\end{split}¥
\end{equation}¥
where $k=\omega/c$, $\omega$ is the frequency of the incident field, $p_{\omega}$ is the dipole strength of the nanoparticle, $\phi_{\omega}$ is the relative phase between the incident field and the induced dipole, $\hat{\vect{z}}$ is the unit vector along the direction of the dipole, and $\hat{\vect{r}}$ is the unit vector in the radial direction from the center of the dipole. $ E_{far}(\vect{r})$, $ E_{mid}(\vect{r})$, and $ E_{near}(\vect{r})$ are the spatial patterns of the far-, intermediate-, and near-fields, respectively. It is clear from \Eq{dipolefield} that the near- and far-fields have the same phase spectrum. Although the power spectra of the near- and far-fields differ by a factor of $\omega^2$, this difference is negligible for a plasmonic resonance curve that is about $100nm$ wide and centered at $800nm$. Thus, the near- and far-fields of a nanoparticle can have an identical temporal pulse shape. The field reflected from the plasmonic antenna is the sum of the fields from many dipole radiators. When the plasmonic antenna is excited by a pulse incident at an angle, the far-fields from all the radiators have the same time delay in the direction specified by the law of reflection. In other directions, the far-fields interfere destructively. Thus, the far-field of the plasmonic antenna has the same temporal shape as the far-field of the individual radiators. The near-field close to the surface of the antenna has contributions from the incident field and the nearest radiators of the antenna. The field in the immediate vicinity of one particular radiator has also a contribution from a neighboring radiator. However that field has a time delay that depends on the distance between the next neighbor pairs. For an array of nanoparticles in an antenna, these distances can be in the subwavelength region. Then the contribution of a next neighbor radiator is delayed in time at most by their distance divided by the speed of light, and therefore, within the sub-cycle oscillation. This contribution would be in phase and not reduce the local field. The near-fields from more distant neighbors can have a significant time delay and may potentially cause the temporal duration of the near-field to broaden. However, the strength of the near-field of any radiator decays quickly over the distance of one wavelength, so the contribution from the far neighbors can be ignored. Consequently, the near-field of the plasmonic antenna has the same temporal shape as the near-field of the individual radiator. Assuming that the antenna is an array of independent dipole radiators, the temporal response of the antenna's far-field equals the temporal response of the antenna's near-field.

\begin{figure}[t]
\centering
\scalebox{0.35}{\includegraphics{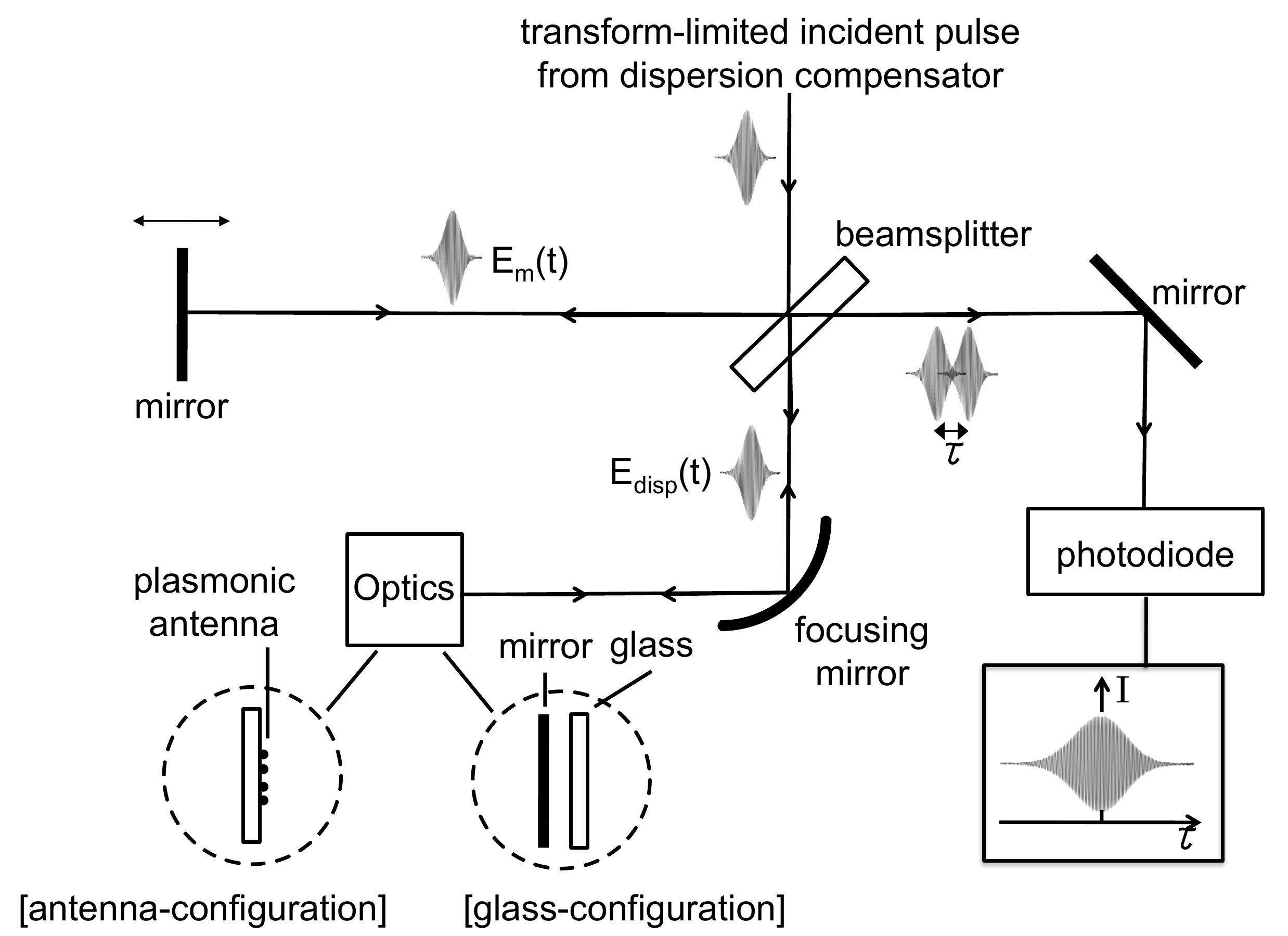}}
\caption{Experimental setups. Experiments are performed for the antenna-configuration and the glass-configuration configuration. The experiment with the glass-configuration tests the validity of the glass-model, while the experiment of the antenna-configuration measures the temporal width of the plasmonic antenna's far field.}
\label{fig:setup}
\end{figure}

Information about the temporal width of the antenna's far-field may be obtained from comparing the cross-correlation signal of the reflected pulse to the auto-correlation signal of the incident pulse. Comparison between the cross-correlation signal and the auto-correlation signal shows the contrast between the reflected pulse and the incident pulse, revealing any broadening that may have occurred. However, without knowing the specific shape of the pulse, its temporal width cannot be inferred directly from the cross-correlation signal. In this study, the cross-correlation signal is modeled by reconstructing the pulse fields from the experimentally obtained power spectrum and the theoretically computed phase spectrum. The phase spectrum of the reflected pulse is modeled by calculating the phase dispersion of a pulse which passes through a glass slide (fused silica) twice,
\begin{equation}
¥	\phi_{disp}(\omega) = \frac{2 \omega d}{c} n(\omega) + \phi_{bs}(\omega),
\end{equation}¥
where $d$ is the thickness of the glass slide, $\omega$ is the frequency, $\phi_{bs}(\omega)$ is the phase dispersion due to the beamsplitter (BK7), and $n(\omega)$ is the index of refraction of the glass (fused silica). The thickness of the glass, $d$, is the only fitting parameter in this ``glass-model'' and is determined by comparing theoretical and experimental cross-correlation signals. To test the validity of this glass-model, experiments were performed in the antenna-configuration and the glass-configuration, shown in \Fig{fig:setup}. After the pulse passed through a dispersion compensator \cite{dispersion}, a beamsplitter splits the pulse to two arms. In the antenna-configuration, the pulse reflected from the plasmonic antenna arm is interfered with the pulse reflected from the mirror arm. The temporal interference pattern is the cross-correlation signal,
\begin{equation}
¥	C(\tau) = \infint E_{m}(t-\tau)E_{disp}(t) \,d\tau,
\end{equation}¥
where $E_{m}(t)$ is the pulse from the mirror arm, and $E_{disp}(t)$ is the pulse from the antenna arm of the interferometer. If the plasmonic antenna changes either the power or phase spectrum of the incident pulse, the pulse from the antenna arm, $E_{disp}(t)$, would have a different shape than the pulse from the mirror arm, $E_{m}(t)$. This would result in a cross-correlation signal between $E_{m}(t)$ and $E_{pl}(t)$ that is broader than the auto-correlation signal of $E_{m}(t)$,
\begin{equation}
¥	A(\tau) = \infint E_{m}(t-\tau)E_{m}(t) \,d\tau.
\end{equation}¥
The same applies to the glass-configuration except that glass causes a known changes in the phase spectrum while leaving the power spectrum unaltered.

\begin{figure}[t]
\centering
\scalebox{0.38}{\includegraphics{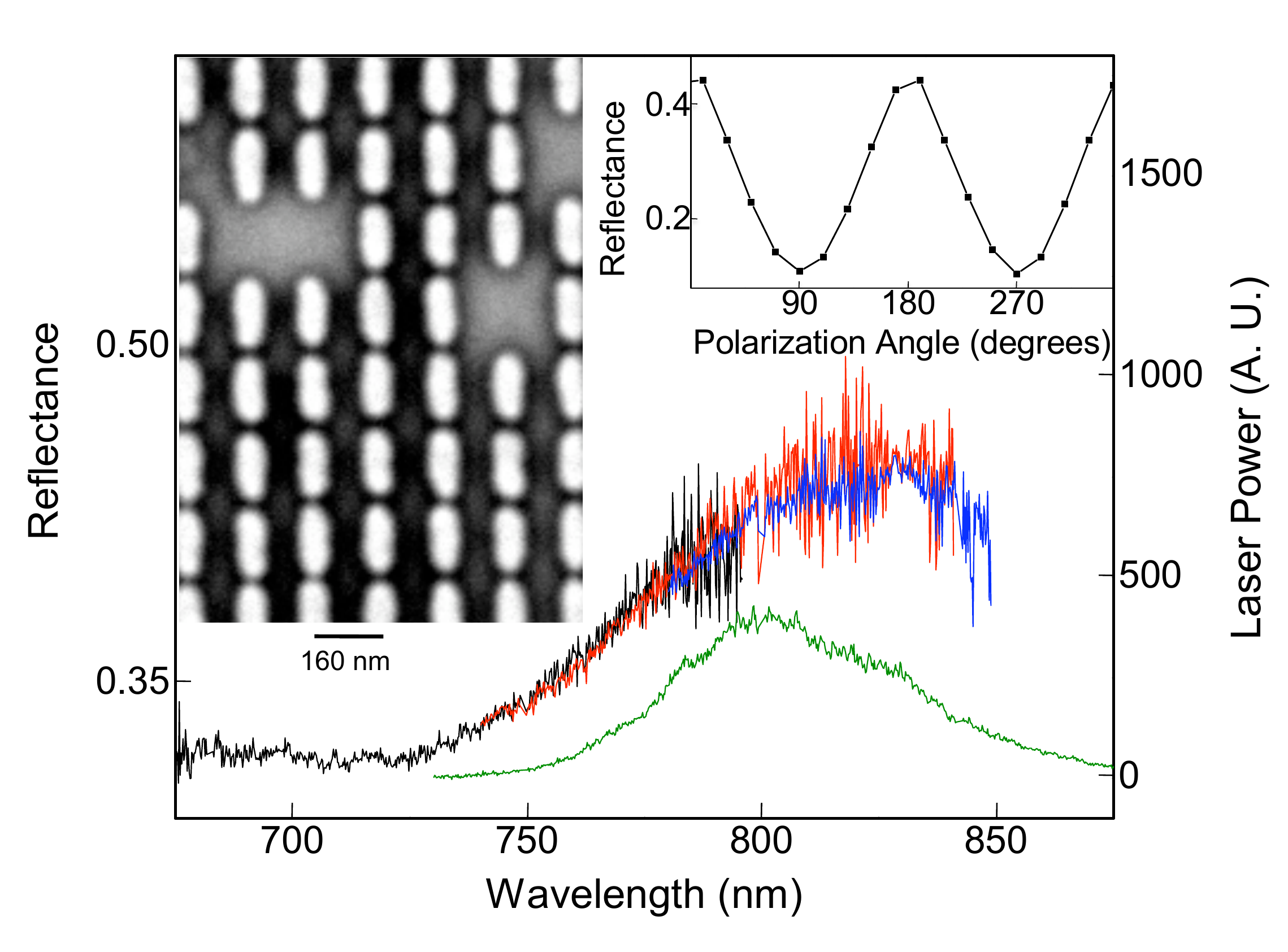}}
\caption{Antenna characterization. Left top: An SEM micrograph of a localized area of the array of gold nanorods that make up the plasmonic antenna.  Missing nanorods can be observed. Main graph: The power spectrum of the input laser pulse (green line) is centered around $800nm$ and supports a minimum pulse duration of $20fs$. The reflectance of incident unpolarized white light from the plasmonic antenna (black, red, and blue lines) shows a broad resonance structure. Right top: The reflectance of linearly polarized light at a wavelength of $800nm$  shows a maximal reflection when the incident polarization is aligned along the length nanorods.}
\label{fig:PAPlot}
\end{figure}

In our experiment, we use a light pulse which has a full-width-half-maximum spectral bandwidth of $\Delta \lambda = 63nm$ and a central frequency of $\lambda_{c} = 800nm$. Assuming a transform-limited pulse shape, the full-width-half-maximum temporal width of the pulse field is $\Delta t \simeq 20fs$. The plasmonic antenna used in the experiment was fabricated using electron-beam lithography as described in \cite{Capasso:resonance}. Images obtained with a scanning-electron microscope (see \Fig{fig:PAPlot}) indicate rod dimensions of $170nm \times 80nm$. The rod array has a period of $160nm$ in the direction perpendicular to the rod length and a spacing of $10-20nm$ along their length.  The entire array has a size of $100 \mu m \times 100 \mu m$.  Defects, such as missing rods and scratches, were present and are likely a result of the fabrication process and/or damage incurred during handling the device.  The broadband reflectance spectrum was measured by focusing white light from a xenon lamp with a $50\times$ microscope objective onto the plasmonic antenna (\Fig{fig:PAPlot}).  Reflected light was collected into a fiber-coupled spectrometer and the resulting spectrum was normalized with the spectrum reflected by a silver mirror.  The result is consistent with the relations between resonance and rod dimensions and spacing reported in \cite{Capasso:resonance}.  Comparison of the input laser spectrum with the plasmonic antenna reflectance spectrum shows that the laser pulse contained frequencies appropriate for generating near-field enhancement in the antenna.  The reflection of the array was also measured using a focused beam of linearly polarized light with a wavelength of $800nm$ (\Fig{fig:PAPlot}).  The polarization-dependence of the reflected power is indicative of the resonance behavior of the antenna.

\begin{figure}[t]
\centering
\scalebox{0.35}{\includegraphics{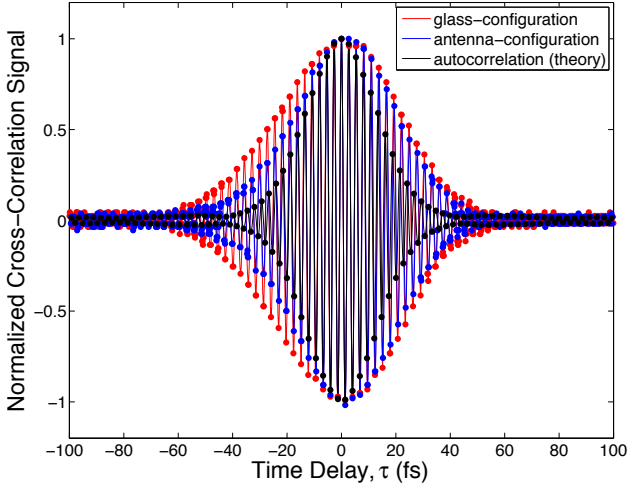}}
\caption{Experimental data of the cross-correlation signals from the antenna-configuration (blue line) and the glass-configuration (red line). The auto-correlation signal of the incident pulse (black line) is computed from the measured power spectrum according to the Wiener-Khinchin theorem.}
\label{fig:cross_correlation}
\end{figure}

\begin{figure}[t]
\centering
\scalebox{0.37}{\includegraphics{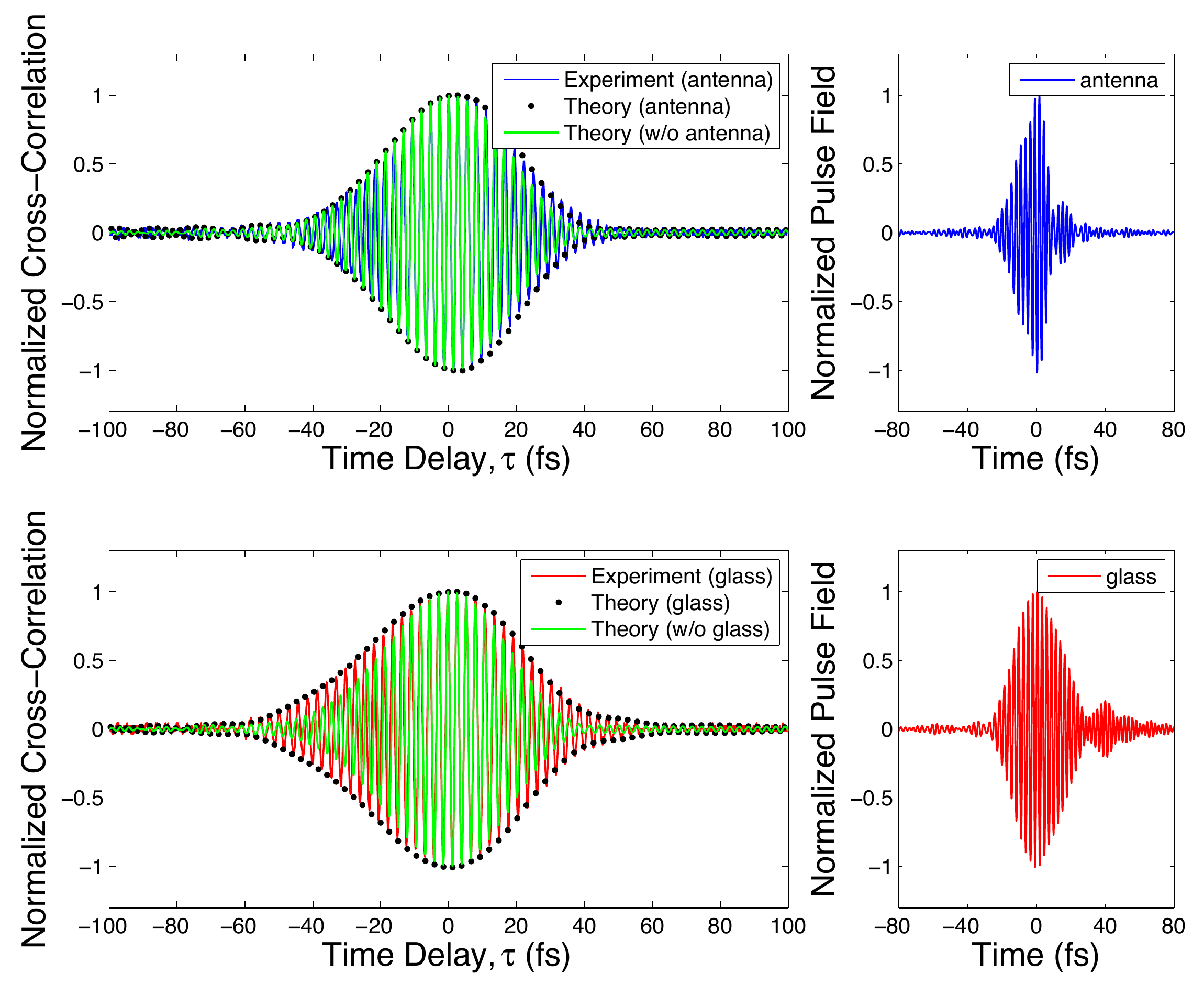}}
\caption{Field reconstruction from the glass-model. Left: The theoretically calculated cross-correlation signal shown in black dots, is fit to the experimental data (blue and red lines) to determine the glass thickness in the glass-model (see text). For comparison the theoretical cross-correlation signal for zero glass thickness is shown (green line). Right: Reflected pulses are reconstructed with the fit parameter obtained from the glass-model in the absence of dispersion from other optical components in the system. The pulse reflected from the array is shown in blue, and the pulse from the glass in red.}
\label{fig:model}
\end{figure}

The experimental results are shown in \Fig{fig:cross_correlation}. The autocorrelation signal is obtained by taking a Fourier transform of the measured power spectrum according to the Wiener-Khinchin theorem. The width of the autocorrelation signal is about $30fs$. The cross-correlation widths measured in the antenna- and glass-configuration are about $40fs$ and $50fs$, respectively. To interpret this result, the glass-model is used. Agreement between the model and the experimental data is found when the fitting parameter is set at $1.145mm$, as shown in \Fig{fig:model}. This is indeed close to the measured thickness of the glass slide used in the experiment, which is $1.14mm$. When this model is applied to the cross-correlation spectrum acquired in the antenna-configuration, the effect of reflection from the antenna can be modeled well by the glass-model when the fitting parameter is set to $0.25mm$. Use of this model enables the reconstruction of the reflected pulse by the nanorod array without inclusion of dispersion from other optical components in the system. The temporal width of the pulse reflected by the antenna (shown in \Fig{fig:model}) is found to be close to that of the incident pulse.

\begin{figure}[t]
\centering
\scalebox{0.44}{\includegraphics{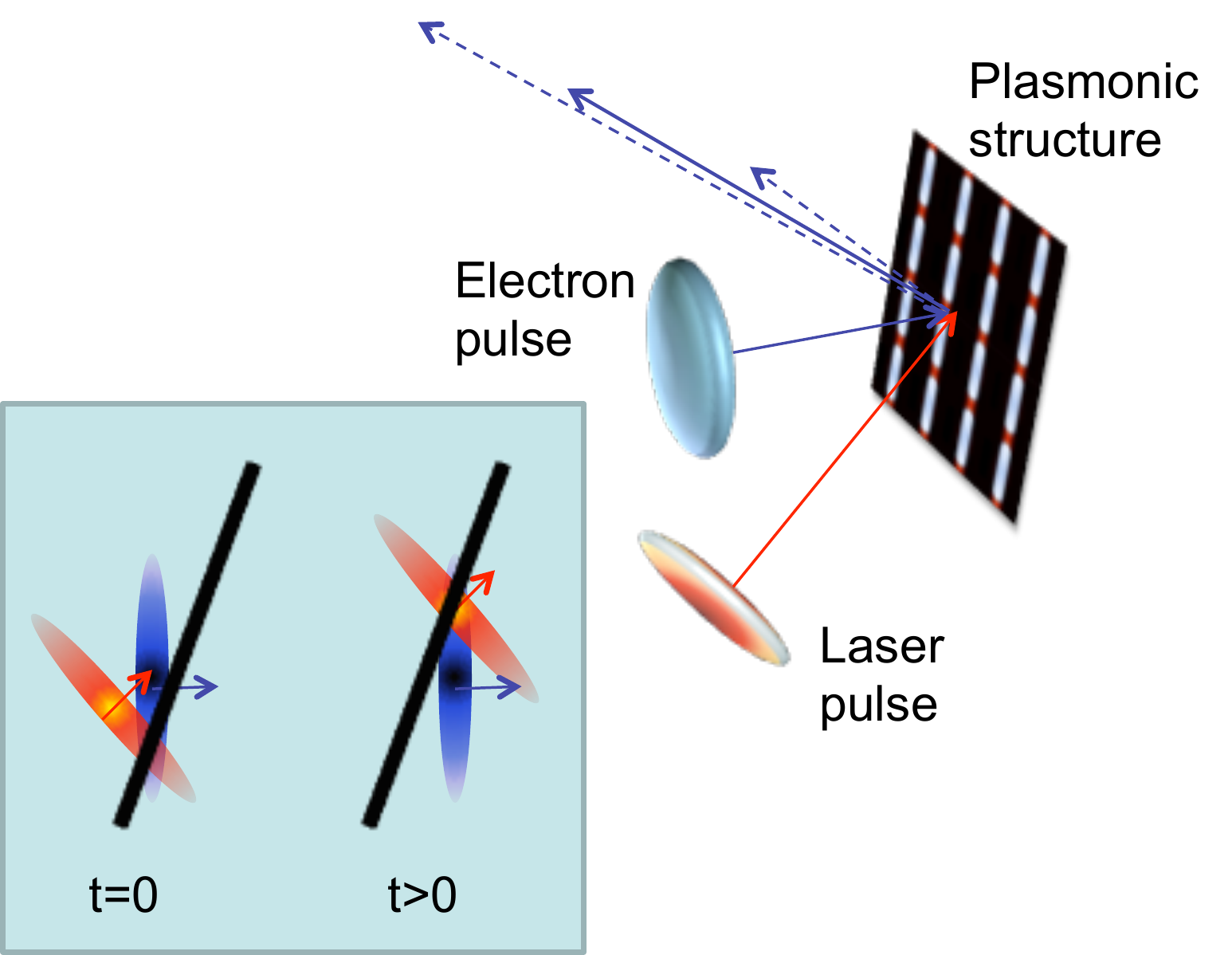}}
\caption{Proposed femtosecond electron switch.  A laser pulse affects the scattering of an electron pulse through the pondermotive potential provided by the enhanced near field of a plasmonic antenna. The intersection of a laser pulse and electron pulse can be directed onto a planar structure. Following an idea of Zewail \cite{Williamson}, the electron velocity and interaction angle can be chosen in such a way that the pulses remain synchronous as they sweep over the planar plasmonic structure. The consequence is that the temporal resolution of the switched electron beams (dashed blue lines) is set by the laser pulse duration.}
\label{fig:femtoswitch}
\end{figure}

Our result indicates that it may be feasible to use the plasmonic antenna as a femtosecond electron switch. Consider an electron pulse that is cross fired with a laser pulse in such a way that the electron pulse and the laser pulse meet at the plasmonic antenna (\Fig{fig:femtoswitch}). Upon excitation by the laser pulse, the antenna will provide a spatially modulated near-field defined by the periodicity of the antenna array. An antenna can enhance the laser intensity at its near field by a factor of $\kappa \simeq 800$ \cite{Capasso:enhancement}, although for the antenna array characterized above an enhancement factor of about $50$ is expected \cite{Capasso:resonance}. As an electron comes close to the antenna surface, it will experience a force from the antenna's near-field. For an electron wave with a small coherence length the interaction time will be shorter than the laser period ($2.7fs$) and the electron will interact with the enhanced electric field of the array. If the electron wave is more delocalized, the electron will experience the cycle-averaged ponderomotive potential of the array \cite{Bucksbaum,Freimund}. The maximum diffraction angle by the ponderomotive potential $\Delta \theta_{m} $ may be estimated by \cite{Batelaan}
\begin{equation}
¥	\Delta \theta_{m} = \frac{\Delta v}{v} = \frac{\tau}{m v} \frac{\Delta U_{p}}{\Delta x} = \frac{\tau}{m v} \frac{e^2 (\kappa I)}{2 m \epsilon_{0} c \omega^2 \Delta x},
\end{equation}¥
where $\tau$ is the interaction time, $v$ is the electron's speed, $m$ is the electron's mass, $\omega$ is the laser frequency, $I$ is the laser intensity, $U_{p}$ is the pondermotive potential of the near-field, and $\Delta x$ is the distance over which the pondermotive potential drops to a small value. For a non-amplified femtosecond laser oscillator operated at $800nm$, the laser pulse has an energy of $10nJ$ and thus an intensity of $I = 2\times10^{11}W/cm^2$ for a duration of $50fs$ and a focus of $10\mu m$. Assuming that the typical interaction length scale between the potential and the electron wave is about $100nm$, the interaction time can be estimated to be $\tau = 10 (fs)$ for a $500eV$ electron pulse. Given an enhancement factor of $\kappa \simeq 800$, a significant diffraction angle of $\Delta \theta_{m} = 10mrad$ can be expected. For a more moderate intensity enhancement of $\kappa=40$, the ponderomotive diffraction is reduced to a measurable $\Delta \theta_{m} = 0.5 (mrad)$. This is considered to be a lower limit for the expected electron diffraction angle.

If the electron wave interaction is limited to the enhanced field within the $10nm$ gap between the nanoparticles, the interaction time will be $1fs$ and classical deflection by the antenna's electric field could be considered. In this case the deflection angle is given by
\begin{equation}
¥	\Delta \theta_{E} = \frac{\Delta v}{v} = \frac{e E\tau}{m v},
\end{equation}¥
where the electric field $E=\sqrt{\kappa \epsilon_{0} c I}$. For a moderate enhancement of $\kappa \simeq 50$ the result is a large deflection of $\Delta \theta_{E} = 50mrad$. These rough estimates indicate that it appears reasonable to consider a plasmonic antenna for the purpose of ultrafast electron switching.

In summary, three properties of a plasmonic antenna justify its proposed use as an ultrafast electron switch. First, the periodic structure of the plasmonic antenna ensures a large gradient of the electric field and thus a large force on the electron. Second, the near-field enhancement eliminates the need for strong laser power, enabling the use of a laser with a higher repetition rate and increasing the number of electrons per second detected. Third, the planar geometry allows for femtosecond resolution detection and control of large diameter electron beams by the use of angle tuning \cite{Williamson}.

In conclusion, agreement between theory and experimental cross-correlation data in the glass-configuration justifies use of the glass-model to model the phase spectrum of the laser pulse reflected from the plasmonic antenna. According to this model, the temporal width of the antenna's far field is shown to be close to that of the excitation pulse. Under the assumption that the plasmonic antenna is described by a collection of dipole radiators, the near field of the plasmonic antenna responds on the femtosecond scale. This ultrafast enhanced near field may affect the motion of incident electron beams. Femtosecond nanoscale pulsed electron sources have been developed \cite{Hommelhoff,Hilbert:1}. Electron pulse compression techniques applied to electron beams extracted from pulsed sources have been shown to deliver electron pulses of about $100fs$ \cite{Oudheusden}. Electron pulse compression techniques have been proposed to deliver sub-femtosecond electron pulses \cite{Baum,Hilbert:2,Veisz}. A plasmonic antenna functioning as an electron switch may be an ideal detection method for these sub-femtosecond electron pulses.

\

W. Huang, M. Becker, and H.Batelaan gratefully acknowledge funding from NSF Grant No.~0969506. The work of E. Smythe and F. Capasso was supported by the Air Force Office of Scientific Research (AFOSR) under a Multidisciplinary University Research Initiative (MURI) Program (Contract FA9550-05-1-0435) and by the Center for Nanoscale Systems (CNS) at Harvard University, a member of the National Nanotechnology Infrastructure Network (NNIN).

\end{document}